\documentstyle[preprint,eqsecnum,aps]{revtex}
\begin{document}
\draft
\title{ Can Schwarzschildean  gravitational fields suppress gravitational
waves?}

\author{Edward Malec}
\address{Institute of Physics, Jagiellonian University,
 {3}0-59 Krak\'ow, Reymonta 4, Poland}

\author{Gerhard Sch\"afer}
\address{Theoretical Physics Institute, Friedrich Schiller University,
07743 Jena, Max-Wien Pl. 1,  Germany}

\maketitle

\begin{abstract}
Gravitational waves  in the linear approximation
propagate in the Schwarz\-schild spacetime similarly as
electromagnetic waves. A fraction of the
radiation scatters off the curvature of the  geometry.
The energy of the  backscattered part of an initially
outgoing pulse of the quadrupole gravitational radiation is estimated
by compact formulas depending on the initial energy, the Schwarzschild
radius, and the location and width of the pulse. The backscatter
becomes negligible in the short wavelength regime.
\end{abstract}
\pacs{ }
\date{ }

\section{ Introduction}

Backscattering has been investigated for a long time
for various wave equations  (see, for instance,
\cite{Hadamard}). In general relativity, this topic has been  studied
since early 1960's (\cite{MTW}, \cite{collective}).
This paper continues the programme that started with the study of the
backscatter of scalar \cite{MENOM} and electromagnetic fields
(\cite{malec2000}
and \cite{meprl}). Here we investigate the propagation of even-parity
gravitational waves in a (fixed) background Schwarzschild spacetime,
assuming  a nonstationary source. The discussion, however, is done without
any reference to the source. We only deal with field quantities.
It is assumed  that initial data are those of
an isolated pulse (burst) of a gravitational wave. The main question
that is answered is what fraction of the initially outgoing radiation
may undergo backscattering before reaching null infinity?
The strength of the backscattering is assessed by bounding the fraction of
the initial burst energy that will not reach a distant observer
in the main pulse.

The even-parity waves are the only waves which are radiated
during the axisymmetric collision of nonspinning black holes \cite{Price},
and since in this case the Schwarzschild spacetime is a valid
starting point for an approximation scheme,
it gives us an opportunity to bound the strength of
the phenomenon in a fairly realistic astrophysical context.

The following five sections of the paper
give a theoretical description of the backscattering effect. The Sec. II
brings notation and the Zerilli equation. In Sec. III, the initial
data are bounded by the initial energy and the solution
is sought in the form of superposition of an outgoing radiation (defined
by initial data) and a backscattered term. The evolution of the
backscattered term can be bounded by solutions of
two differential inequalities. The bounds that are derived in Sec. IV and
in the Appendix deal with a general situation; no assumption is made
about the initial radiation. In Sec. V, we discuss initial data that are
of compact support and in addition, the relative width of the support is
small. Such data correspond to radiation that is dominated
by short wavelengths. In this case stronger estimates are derived.
They imply that in the limit of short wavelengths (relative
width of the support tending to zero) the backscattering effect
becomes negligible. In Sec. VI, the "small relative width condition"
of Sec. V  is supplemented by the assumption that the initial
burst is far away from the horizon.

\section{Formalism}

The space-time geometry  is defined  by
a Schwarzschildean  line element,
\begin{equation} ds^2 = - (1-{2m\over R})dt^2 +
{1\over 1-{2m\over R}} dR^2 +
R^2 d\Omega^2~,
\label{1}
\end{equation}
where $t$ is a time coordinate, $R$ is a radial
coordinate that coincides with the areal radius
and $d\Omega^2 = d\theta^2 + \sin^2\theta d\phi^2$
is the line element on the unit sphere, $0\le \phi < 2\pi $
and $0\le \theta \le \pi $.
Throughout this paper $G$, the Newtonian gravitational constant,
and $c$, the velocity of light are put equal to 1.

As explained in the Introdction, we restrict ourselves,
 to the even-parity axial perturbations.
Their propagation is
ruled, in the linear approximation,  by the Zerilli equation
\cite{Zerilli}.
Formulated in terms of the gauge-invariant amplitude $\Psi $
 defined by Moncrief
\cite{moncrief}, this equation reads, in the case of $l=2$ multipole
 \cite{Price},
\begin{equation}
(-\partial_t^2 + \partial_{r^*}^2)\Psi = V\Psi ,
\label{2}
\end{equation}
where  the potential $V$ is given by
\begin{equation}
V(R)=6(1-{2m\over R})^2  {1\over R^2} +
(1-{2m\over R}){63m^2(1+{m\over R})\over 2R^4(1+{3m\over 2R})^2},
\label{3}
\end{equation}
and where
\begin{equation}
r^*=R+2m\ln ({R\over 2m}-1)
\end{equation}
is the tortoise radial coordinate.

Consider a set of functions $ \Psi_i(r^*-t)$, $i=0,1,2$,
that satisfy the following  linear relations
\begin{eqnarray}
&&\partial_t  \Psi_1= 3  \Psi_0,\nonumber\\ &&
\partial_t   \Psi_2= \Psi_1 - m\partial_t \Psi_1.
\label{4}
\end{eqnarray}
The combination
\begin{equation}
\tilde \Psi \equiv  \Psi_0(r^*-t)+ {  \Psi_1(r^*-t)\over R} +
{  \Psi_2(r^*-t)\over R^2}
\label{4a}
\end{equation}
 solves Eq. (\ref{2}) in  Minkowski space-time ($m=0$);
it represents purely outgoing radiation.

 Let the initial data
of a  solution $\Psi $ of (\ref{2}) coincide with  $\tilde \Psi $
at $t=0$.
Then, initially, $\Psi $ represents a purely outgoing  wave.
It should be noted that the assumption   that initial data are
(initially) purely  outgoing is made in this paper  only for the
sake of clear presentation. In the linear approximation
the propagation of the  initially outgoing radiation
is independent of whether or not   ingoing radiation is present.

We decompose  the sought  solution $\Psi (r^*,t)$
into the known part $\tilde \Psi $ and an unknown function $\delta $
\begin{equation}
\Psi =\tilde \Psi +\delta .
\label{5}
\end{equation}
Due to the choice of  the initial data  made above one has
$\delta =\partial_t\delta =0$, at $t=0$.

\section{Energy estimates}

Let us assume that the  quadrupole  initial data are
 defined by a smooth triad of the
functions $\Psi_k$ ($k=0,1,2$)  with the initial  support  $[a, b]$
 ($b< \infty $).
That  guarantees that the initial  energy density multiplied by $R^2$,
$\rho =\bigl( (\partial_t\Psi )^2 + (\partial_{r^*}\Psi )^2
+V\Psi^2\bigr) /\eta_R$,  is smooth and
vanishes on the boundary $a$.   Here $\eta_R=1-2m/R$ holds.

The energy content inside  a part of a  Cauchy hypersurface $\Sigma_t$
that is exterior to a ball of a radius $R$ can be defined as
$E(R,t)\equiv \int_{R}^{\infty }dr \rho (r,t)$.
Let us point out that in order to ensure a proper normalization
of the energy flux at infinity, there should be a normalization constant
in the definition of the energy $E(R,t)$. We decided to omit it,
since later on we shall  be interested only in the relative efficiency
of the backscatter and thus the normalization factor cancels out.
The total initial energy corresponding to the  hitherto defined
initial data is equal to $E(a,0)$.

{\bf Lemma 1.} Defining
\begin{eqnarray}
&&C_1\equiv {3\over 2}\sqrt{ (2+\sqrt{2/3})E(a,0) } ,\nonumber\\
&& C_2\equiv  \sqrt{2E(a,0)  }     ,
\nonumber\\
 &&C_3\equiv \sqrt{2E(a,0) \over \eta_a(2\sqrt{6}+1)},
\label{6}
\end{eqnarray}
and introducing the  two nonnegative functions
\begin{eqnarray}
g_1(R)& =&\ln \Biggl( {-2 m + R \over a-2 m }\Biggr)
+ 32 m^5\Biggl( {-1\over 5(-2m +
R)^5}+{1\over 5(-2 m + a)^5}\Biggr)
+\nonumber\\
&&20m^4\Biggl( {-1\over (-2 m + R)^4} +{ 1\over (-2 m + a)^4}\Biggr)
+80m^3\Biggl( {-1\over 3 (-2 m + R)^3} + {1\over 3 (-2 m + a)^3}\Biggr) +
\nonumber\\
&& 20 m^2\Biggl( {-1\over (-2 m + R)^2} + {1\over (-2 m + a)^2}\Biggr)
+10m\Biggl(   {-1\over  -2 m + R}+ {1\over -2 m + a}\Biggr)
\label{6a}
\end{eqnarray}
and
\begin{eqnarray}
 g_2(R) &=&R-a +
16 m^4\Biggl( {-1\over 3 (-2 m + R)^3}
+{1\over 3 (-2 m + a)^3}\Biggr) +\nonumber\\
&& (16 m^3)\Biggl({-1\over (-2 m + R)^2} +{1\over (-2 m + a)^2}\Biggr)
+ \nonumber\\
&& (24 m^2)\Biggl( {-1\over -2 m + R} +{1\over -2 m + a}\Biggr) +
 8 m \ln \Biggl( {-2 m + R\over -2 m+a}\Biggr)             ,
\label{6b}
\end{eqnarray}
   the  following inequalities hold  at $t=0$ and for $R\ge a$:
\begin{eqnarray}
&&{|\Psi_1(R)|\over R^{3/2}}\le C_1 \eta_R^{3/2}\sqrt{ g_1(R) } ,
  \nonumber\\
&& {| \Psi_2(R)|\over \sqrt{\eta_R}R^{2}}\le C_2   \sqrt{g_2(R)} +
C_1{6m\over \sqrt{a}}  \sqrt{g_1(R)}\Bigl( 1-\sqrt{a\over R}\Bigr)
 + \nonumber \\
&&6C_3{m\over \sqrt{a}}
\sqrt{1-\bigl( {a\over R}\bigr)^{2\sqrt{6}+1}}\Biggl( {1\over
 \sqrt{\eta_a}}
-  {1\over \sqrt{{R\over a}-{2m\over a}}} \Biggr)   ,
\nonumber \\
 &&{|\tilde \Psi (R)|\over \sqrt{R}} \le  C_3
\sqrt{1-\bigl( {a\over R}\bigr)^{2\sqrt{6}+1}},\nonumber\\
&&|{\Psi_0(R)|\over  \sqrt{R}} \le
C_3\sqrt{1-\bigl( {a\over R}\bigr)^{2\sqrt{6}+1}}+ \nonumber \\
&&  C_1 \eta_R^{3/2}\sqrt{ g_1(R) } +    C_2   \sqrt{g_2(R)\over R} +
C_1{6m\over \sqrt{aR}}  \sqrt{g_1(R)}(1-\sqrt{a\over R}) + \nonumber \\
&&6C_3{m\over \sqrt{aR}}
\sqrt{1-\bigl( {a\over R}\bigr)^{2\sqrt{6}+1}}\Biggl( {1\over
 \sqrt{ \eta_a}}
-  {1\over \sqrt{{R\over a}-{2m\over a}}}\Biggr) .
\label{7}
\end{eqnarray}
{\bf Proof.}

One can explicitly verify that
\begin{eqnarray}
&&-\eta_R\Biggl( {   \Psi_1\over R^2} + 2{ \Psi_2\over R^3}\Biggr)=
\partial_t  \tilde \Psi +\partial_{r^*}\tilde \Psi ,\nonumber\\
&&\eta_R\Biggl(  2{ \Psi_0\over R}+{  \Psi_1\over R^2} \Biggr) =
\partial_t\tilde \Psi +\partial_{r^*} \tilde \Psi +{2\eta_R\tilde
\Psi\over R}.
\label{8}
\end{eqnarray}
The equations  (\ref{8}), using the relations
(\ref{4a}), result in
\begin{eqnarray}
&&\partial_R\Bigl( \eta_R^{-3/2}{\Psi_1\over R^{3/2}}\Bigr)
=-{3\over 2R^{1/2}\eta_R^3}
\Biggl( {\partial_t\tilde \Psi \over \eta_R^{1/2}}+
\sqrt{\eta_R}\partial_R\tilde \Psi +
{2\tilde \Psi \sqrt{\eta_R}\over R}\Biggr) ,\nonumber\\
&& \partial_R\Bigl( \eta_R^{-1/2}{\Psi_2\over R^{2}}\Bigr) =
{1\over  \eta_R^2}
\Biggl( {\partial_t\tilde \Psi \over \eta_{_R}^{1/2}} +
\sqrt{\eta_{R}}\partial_R\tilde \Psi \Biggr)
+{3m\over R\eta_{_R}^{3/2}}\Biggl( {\tilde \Psi \over R}-{\Psi_1\over
 R^2}\Biggr) .
\label{8a}
\end{eqnarray}
The integration from $a$ to $R$ and  the use of the Schwarz
inequality yields
\begin{eqnarray}
{|\Psi_1(R)|\over R^{3/2}}\le {3\eta_R^{3/2}\over 2}
\sqrt{(2+\sqrt{2/3})E(a,0) }\Biggl( \int_a^Rdr
{1\over r\eta_r^6}\Biggr)^{1/2}.
\label{8b}
\end{eqnarray}
 Integrating $\int_a^Rdr {1\over r\eta_r^6}$, one immediately arrives at
the first of the postulated  inequalities.

In order to show the third inequality,
notice that $|\tilde \Psi (R)|R^{\sqrt{6}}=
|\int_a^Rdr\partial_r(\tilde \Psi (r)r^{\sqrt{6}})|$.
The latter expression  is bounded from  above, using the Schwarz
inequality, by
\begin{eqnarray}
&&\sqrt{2\int_a^Rdr \Bigl( \eta_r(\partial_r\tilde \Psi )^2+
6\eta_r\tilde \Psi^2/r^2\Bigr) }
\sqrt{\int_a^Rdr
\eta^{-1}_rr^{2\sqrt{6}}} \le \nonumber\\
&& \sqrt{2E(a,0) \over \eta_a (2\sqrt{6}+1)}
R^{\sqrt{6}+0.5} \Biggl( 1-\Bigl( {a\over R}\Bigr)^{\sqrt{6}+1}
\Biggr)^{1/2} ,
\label{3.7}
\end{eqnarray}
where the inequality in Eq. (\ref{3.7}) follows from the monotonicity
of the energy as function of $R$.
The first factor on the left hand side of this inequality is not greater
than $\sqrt{2E(a,0) }$ since $6\eta_r /r^2\le V(r)/\eta_r$.
The  replacement of $\eta_r^{-1}$ by $\eta_a^{-1}$ and the
integration of the other factor leads to   the desired result.

The second of the  equations  of (\ref{8a}) can be integrated.  The
Schwarz inequality  and direct integration as well as  the
 $\tilde \Psi $ and $\Psi_1 $  estimates
should be used in order to get the second inequality of Lemma 1.
The $\Psi_0$-estimate, in turn,
 follows from the identity $\Psi_0=\tilde \Psi
-\Psi_1/R-\Psi_2/R^2$ and the preceding estimates.

\section{The estimate of the diffused  energy}

Let us define the  strength
 of the backscattered   radiation that is directed inward   by
\begin{equation}
h_-(R,t) ={1\over \eta_R }(\partial_t+\partial_{r^*})\delta (R,t).
\label{20}
\end{equation}
Let the outgoing null geodesic  $\tilde \Gamma_{(R,t)} $
  originate at $(R,t)$. If a  point lies on the initial
hypersurface, then we will write
$\tilde \Gamma_{(R,0)}\equiv  \tilde \Gamma_R$.
By $\tilde \Gamma_{(R_0, t_0), (R,t)}$ will be understood a segment of
 $\tilde \Gamma_{(R_0, t_0)}$ ending at $(R,t)$.

A straightforward calculation shows that the rate  of the energy change
along $\tilde \Gamma_{a}$ is given by
\begin{eqnarray}
&&(\partial_t+\partial_{r^*})E(R,t)= \nonumber\\
&&- \Biggl[ \eta_R^2 h_-^2(R,t)
  +V \delta^2(a) \Biggl] .
\label{28}
\end{eqnarray}
It is necessary to point out that in the case of the initial point $R_0>a$
the result would be more complicated; the differentiation of the energy
along $\tilde \Gamma_{R_0}$
would depend also on $\Psi_0, \Psi_1$ and $\Psi_2$. If, however, the
outgoing null geodesics is $\tilde \Gamma_{a}$, then it starts from $a$
where   $\Psi_0, \Psi_1$ and $\Psi_2$  do vanish. Since these functions
depend on the difference $r^*-t$, their values along outgoing geodesics
are constant, and that  allows one to conclude that they vanish at
 $\tilde \Gamma_a$.

The  energy loss, that is the amount of energy that diffused
inward $\tilde \Gamma_a$ is equal to a line integral  along     $\tilde
\Gamma_a$,
\begin{eqnarray}
&& \delta E_a \equiv  E(a,0)- E_{\infty }=
 \nonumber\\ &&
 \int_{a}^{\infty  } dr
  \Biggl[ \eta_r h^2_-
  +{ V \delta^2\over \eta_r } \Biggl] .
\label{29}
\end{eqnarray}
Our goal is to find an estimate of  $\delta E_a $ of a single pulse of
radiation basing only on the information about the position and the energy
of  the initial pulse. Obviously, $0\le \delta E_a \le E(a,0) $ holds.
We are interested in deriving in this section a frequency-independent
bound,
 but later we  obtain  estimates that  are frequency-sensitive.

$\delta $ is initially  zero and its evolution is governed by the
following equation
\begin{eqnarray}
&&(-\partial_t^2 + \partial_{r^*}^2)\delta  = V\delta +
(V-6{\eta_R^2\over R^2})\Biggl( \Psi_0+{\Psi_1\over R}+{\Psi_2\over
R^2}\Biggr) +
\nonumber\\
&&  { 2m\eta_R\over R^4}\Biggl[ -3\Psi_1 +2{\Psi_2\over R}  \Biggr] .
\label{11}
\end{eqnarray}
 One can  define  an  "energy"  $H(R,t)$ of the  field $\delta$  which is
contained in the exterior of a sphere of  radius $R$ as follows
\begin{equation}
H(R,t) = \int_{R}^{\infty }dr
\Bigl(  {(\partial_t\delta )^2\over \eta_r} + \eta_r
(\partial_r\delta)^2+\delta^2{ V\over \eta_r}\Bigr) .
\label{14}
\end{equation}
The rate of change of $H$ along $\tilde \Gamma_{(R,t)} $
is given by
\begin{eqnarray}
&&
(\partial_t+\partial_r^*)H(R,t)= \nonumber\\ &&
-\eta_R\Biggl[ \eta_R
\Bigl( {\partial_t\delta \over \eta_R} +
\partial_R\delta
 \Bigr)^2 +{ V\over  \eta_R}\delta^2 \Biggr] - \nonumber\\
&&4m \int_R^{\infty }dr\eta_r{\partial_t\delta  \over r^4 }\Biggl[
 -3 \Psi_1 +{2\Psi_2\over r}   +
{ 63m (1+{m\over r})\over 4 (1+{3m\over 2r})^2}\Biggl(  \Psi_0+{\Psi_1
\over r}
 +{\Psi_2\over r^2}\Biggr) \Biggr]
 \le
\nonumber\\ &&
4m \int_R^{\infty }dr {\partial_t\delta \over r^4 }\Biggl[
-3 \Psi_1 +{2\Psi_2\over r}   +
{ 63m (1+{m\over r}) \over 4 (1+{3m\over 2r})^2}\Biggl(  \Psi_0+{\Psi_1
\over r}
 +{\Psi_2\over r^2}\Biggr)
  \Biggr] .
\label{15}
\end{eqnarray}
Herein
the inequality follows from the omission of the nonpositive boundary term.
This allows one to estimate the maximal value $H_M$ of the $\delta
$-energy $H$, namely
\begin{equation}
\sqrt{H_M} \le 10.43{m\sqrt{E(a,0) }\over a}+O(m^2).
\label{15.0}
\end{equation}
The calculation is essentially simple, but the  algebra is quite lengthy
and some numerical integrations are required.  Details are relegated to
the  Appendix.  We would like to point out that the $O(m^2)$ terms
  become dominant only
 when the location of the initial radiation pulse is smaller than $6.6m$.
 At $a=15m$  the neglected terms contribute much less than the leading
   term proportional to $m$.

 Now, the integration of the first part of Eq. (\ref{15}) along
 $\tilde \Gamma_{(a,0)} $ yields
\begin{eqnarray}
&&H(\infty )-H(0 )=\nonumber \\
&& -\int_a^{\infty } dR \Biggl[ \eta_R
\Bigl( {\partial_t\delta \over \eta_R} +
\partial_R\delta\Bigr)^2 +{ V\over  \eta_R}\delta^2 \Biggr] - \nonumber\\
&&\int_a^{\infty }dR
4m \int_R^{\infty }dr {\partial_t\delta \over r^4 }\Biggl[
-3 \Psi_1 +{2\Psi_2\over r}   +
{ 63m (1+{m\over r}) \over 4 (1+{3m\over 2r})^2}\Biggl(  \Psi_0+{\Psi_1
\over r}
 +{\Psi_2\over r^2}\Biggr)
 \Biggr].
\label{e1}
\end{eqnarray}
Initially,  $H$ vanishes (both $\delta $ and $\partial_t\delta $ vanish)
and $H$ is manifestly nonnegative.    The first integral on the right
hand
side  of (\ref{e1}) is recognized to be just $\delta E_a$. The second
integral
in turn can be shown to be bounded -   using the Schwarz inequality and
then the results  of the Appendix - by $2\sqrt{H_M}
(10.43{m\sqrt{E(a,0) }\over a}+O(m^2))$.   Thus,   (\ref{e1}) implies
\begin{equation}
\delta E_a \le 2\Bigl( 10.43{m\sqrt{E(a,0) }\over a}+O(m^2)\Bigr)
\sqrt{H_M} \le
\Biggl[ 54.5 \Biggl( {2m\over a}\Biggr)^2 +O(m^3)\Biggr] E(a,0) ;
\label{e2}
\end{equation}
the right hand side of the first inequality achieves a
 maximal value when $H$ is maximal and that implies the second inequality.

 Thus in summary,  for the  fraction
 of the   energy that could diffuse through the null cone $C_a$,
 it holds:

{\bf Theorem.}      $\delta E_a/E(a,0)  $
satisfies the inequality
\begin{eqnarray}
&&  {\delta E_a\over E(a,0)} \le  54.5 \times \Bigl( {2m\over a}\Bigr)^2
  +O(m^3/a^3).
\label{27}
\end{eqnarray}
We would like to point out that the above derivation is more efficient and
simpler than the one used in   \cite{malec2000}   or   \cite{meprl}
 when $\delta E_a$ was
estimated directly on the basis of the estimates of $\delta $ and $h_-$.
This  alternative approach would  require a laborious integration of the
field equation and the final estimate would be much worse.

\section{The wavelength  of the initial radiation and the backscatter.}

In this section we shall consider the backscatter of the radiation
that is initially of compact support and, in addition, the condition
$(a-b)/a<<1$ is satisfied. The leading contribution
 - only terms that are quadratic in $m^2$ -  will be  found.

Under the above conditions one infers from  Eq. (\ref{7})
that on the initial hypersurface
\begin{eqnarray}
|\Psi_1(R)|b^{3/2}\le C_1  b^{3/2}\sqrt{ g_1(R) }\le
C_1\sqrt{b-a\over a} b^{3/2}
\label{5.1}
\end{eqnarray}
and
\begin{equation}
 | \Psi_2(R)|    \le C_2  b^{2} \sqrt{b-a}
\label{5.2}
\end{equation}
are valid. With the same accuracy the inequality (\ref{15b}) of the
Appendix  reads
\begin{eqnarray}
&&
(\partial_t+\partial_{r^*})\sqrt{H(R,t)}
 \le
2m\Biggl( \int_R^{ R(b) }dr\eta_r {9\Psi_1^2\over r^8}\Biggr)^{1/2}+
2m\Biggl( \int_R^{\infty }dr{4\Psi_2^2\over r^{10}} \Biggr)^{1/2}  \le
\nonumber\\
&& 6mC_1\sqrt{b-a\over a} b^{3/2}\Biggl(
 \int_R^{R(b)  }dr{1 \over r^8}\Biggr)^{1/2}+
4mC_2  b^{2} \sqrt{b-a}\Biggl( \int_R^{ R(b)}dr{ 1\over r^{10}}
\Biggr)^{1/2}.
\label{5.3}
\end{eqnarray}
Herein, the integration extends from $R$, where $R\in \tilde \Gamma_a$,
to $R(b)$, which is defined by  $(R(b),t)\in \tilde \Gamma_b$.
One has $R(b)-R=b-a$ up to the term $m^0$.
The  integral $\int_R^{R(b)  }dr{1 \over r^8}$ is bounded from above
by $(b-a)/R^8$ and the integral $\int_R^{ R(b)}dr{ 1\over r^{10}}$
is bounded from above by   $(b-a)/R^{10}$, again to lowest order in
powers of $m$.

Thus  one arrives at
\begin{eqnarray}
(\partial_t+\partial_{r^*})\sqrt{H(R,t)} \le
4m(b-a)b^{3/2}\Biggl( { 1.5C_1 \over \sqrt{a} R^4}   +  {C_2\sqrt{b}\over
R^5 }\Biggr) .
\label{5.4}
\end{eqnarray}
The integration of this inequality along the null geodesic $\tilde
\Gamma_a $
yields
\begin{eqnarray}
\sqrt{H_M }\le   m(b-a){b^{3/2}\over a^{7/2}}
\Bigl(  2C_1     +   C_2 \Bigr) +O(m^2) .
\label{5.5}
\end{eqnarray}
Taking into account the condition that $b-a<<a$, one arrives at
\begin{eqnarray}
H_M\le   {4m^2\over a^2}\Biggl( {b-a\over a}\Biggr)^2
\Biggl( {b\over a}\Biggr)^3
\Biggl(  C_1     +  {C_2\over   2 }\Biggr)^2 +O(m^3/b^3) .
\label{5.6}
\end{eqnarray}
Since the amount of backscattered energy $\delta E_a$ is bounded from
above by  $2H_M$, as shown in the preceding section,
 one finally arrives at the following estimate
\begin{eqnarray}
&&  {\delta E_a\over E(a,0)} \le {8m^2\over a^2}
\Biggl( {b-a\over a}\Biggr)^2  \times    \Biggl( {b\over a}\Biggr)^3
\Biggl(  {3\over 2}\sqrt{2+\sqrt{2\over 3}}    +  { 1\over  \sqrt{ 2 }}
\Biggr)^2 +O(m^3/a^3)\le \nonumber \\
&&  84     {m^2\over a^2}  \Biggl( {b\over a}\Biggr)^3
\Biggl( {b-a\over a}\Biggr)^2 +O(m^3/a^3) .
\label{5.7}
\end{eqnarray}
If $(b-a)/a<0.1$, then the above formula predicts
\begin{eqnarray}
  {\delta E_a\over E(a,0)} \le
  0.84   {m^2\over a^2}   .
\label{5.8}
\end{eqnarray}
It is clear that if the relative width of the initial pulse  tends to zero
then the effect becomes negligible. This can be translated
 into the dependence on the wavelength of the radiation \cite{meprl}:
The compression of the support of a function leads to the
 decrease of its wavelength scale in its Fourier transform.

A careful analysis of the higher order terms would show that they give
a contribution to (\ref{5.7}) that  also scales like
$\Biggl( {b-a\over a}\Biggr)^2$.   In the case when $a\approx 2m$,
Eq. (\ref{5.7}) would be of the form
\begin{equation}
{\delta E_a\over E(a,0)}
 \le C( x)   \Biggl( {b-a\over 2m}\Biggr)^2,
\label{5.9}
\end{equation}
where $C(x)$ is a large number and $x\equiv 2m/a$.
One can show that $\lim_{x\rightarrow  1 }
C(x) =\infty $, but on the other hand $C(x)$ is  fixed, when $2m/a$ is
fixed.
Thus (\ref{5.9})  implies that when $b\rightarrow a$, then the backscatter
  becomes negligible. Radiation that is  dominated  by
  infinitely short wavelengths does not backscatter.

\section{More estimates on high frequency radiation}

We assume  initial data  of compact support $[a,b]$.
 The   initial energy $E(a,0)$ (see the beginning of Sec. III) reads,
 expressed in
terms of functions $  \Psi_0,\Psi_1$ and $\Psi_2$, as follows
\begin{eqnarray}
E(a,0)&=&\int_a^bdr\rho =  \int_a^bdr\Biggr[
   {6  \Biggl( r \bigl( r \Psi_0 (r) +\Psi_1 (r) \bigr)  +
     \Psi_2 (r) \Biggr)^2\over r^6}  +
\Biggl(  \Psi_0 '(r) + {r  \Psi_1 '(r) +
   \Psi_2 '(r) \over r^2} \Biggr)^2 + \nonumber    \\
&&
 { \Biggl( -2 \Psi_2 (r)  - r \bigl( \Psi_1 (r) +
  r ( r \Psi_0 '(r) +  \Psi_1 '(r))  +  \Psi_2 '(r)
   \bigr)  \Biggr)^2\over r^6}\Biggr] .
\label{6.1}
\end{eqnarray}
The radiation energy in the wave zone is known to be
$E(a,0)= C\int_a^bdr(\Psi_0')^2$.
This can be compatible with (\ref{6.1})
(modulo a normalization constant, which is not relevant here), if
 the terms    with $(\Psi_0')^2$ give a leading contribution.

One  notices,  that if $\Psi_{\mu }(R)$ ($\mu =0,1,2$) are
of compact support, then
$|\Psi_{\mu }(R)|=|\int_a^Rdr\partial_r\Psi_{\mu }(r)|\le \sqrt{(R-a)
\int_a^Rdr\partial_r\Psi_{\mu }^2(r)}$. Combining this with (\ref{4})
one arrives  at
\begin{eqnarray}
&&3|\Psi_0(r)|= |\Psi_1'(r)|\le 3\sqrt{b-a}\sqrt{\int_a^bdr(\Psi_0')^2},
\nonumber\\
&&
|\Psi_1(r)|= |\Psi_2'(r)|\le 2 (b-a)^{3/2}\sqrt{\int_a^bdr(\Psi_0')^2},
\nonumber\\
&& |\Psi_2(r)| \le 0.8 (b-a)^{5/2}\sqrt{\int_a^bdr(\Psi_0')^2}.
\label{6.2}
\end{eqnarray}
Taking into account (\ref{6.2}), one concludes that   if
\begin{equation}
 C_1 (b-a)/a <<1,
\label{6.3}
\end{equation}
($C_1$ is a constant of the order of 100)
then the energy is well approximated by $E(a,0)= 2\int_a^bdr(\Psi_0')^2$.

In such circumstances it is clear that our analysis can be greatly
simplified.
First of all the contribution coming from $\Psi_2$ to the backscatter
 is much smaller than that due to $\Psi_1$; notice an additional power of
$(b-a)/a$ in the relevant estimate of (\ref{6.2}). Secondly, the
estimate (\ref{5.1}) of $\Psi_1$ is now replaced by a stronger  result
\begin{eqnarray}
 |\Psi_0(r)| \le  {1\over \sqrt{2}} (b-a)^{3/2}\sqrt{E(a,0) }.
\label{6.4}
\end{eqnarray}
The repetition of the calculation of Sec. V gives finally (taking into
account the above conditions)
\begin{eqnarray}
  {\delta E_a\over E(a,0)} \le \Biggl({2m\over a}\Biggr)^2
\Biggl( {b-a\over a}\Biggr)^4.
\label{6.5}
\end{eqnarray}

\section{Conclusions}

In our paper we derived upper bounds for the backscattering of
gravitational quadrupole waves propagating outward from a central compact
object.
The calculations were restricted to situations where the initial
configuration
was either an arbitrarily shaped wave with support outside some radius $a$
or the wave was a sharp pulse, i. e. its extension was small compared
to its initial location $a$. The obtained upper bounds show that,
for a given central object, the backscattering is the weaker the more
outside from the central object the waves start propagating, and that is
also the weaker the more compact the pulses are, i. e. the higher the
involved frequencies
are. The both results do confirm previous completely different
calculations by
Price, Pullin and Kundu \cite{Kundu}. Backscattering should thus be
strongest
for pulses which start propagating outward close to the horizon of a
black hole. This claim, however, needs further investigation for the
following reasons. First, we gave bounds from above and not from
 below for the amount of backscattered energy.  Second,
the linear approximation may not be accurate enough very close to the
horizon.

Results of Flanagan  and Hughes \cite{Flanagan} and
Buonanno and Damour \cite{Damour} have shown that
the merger part of the gravitational wave
signal could be a significant part of the total energy emitted.   The
wave pulse during the merger phase can be inside 3m.    For a very compact
pulse
located in this region the  inequality (\ref{5.9})
of Sec. V can  still yield a nontrivial bound,
but  in the general case our estimates fail.  The main reason
why we are loosing much in the accuracy is that we
are forced to use several times - for the sake of generality -
  the Schwarz inequality.
The present bounds can   be  significantly improved  if
  initial data  are explicitly known, since in this case
they can  be numerically bounded by an exact expression involving
the initial energy and
the Schwarz inequality would be used only once.
On the other hand, it has been discovered that
the backscattering can be quite strong when
a signal propagates from within the photon sphere \cite{Karkowski}.

In a forthcoming  paper we shall discuss, and compare with the
results of our present paper, several aspects of the
backscattering of gravitational waves where the sources of the
gravitational waves are taken into account.

{\bf Acknowledgements.}  This work has been suported in part  by
the KBN grant 2 PO3B 010 16. One of the authors (EM) gratefully
acknowledges
financial support from the DAAD during his visit in Jena.

\section{Appendix}

In order to show the estimate (\ref{15.0}) one begins with the second
inequality of (\ref{15}).
Notice that $H(t=0)=0$, since $\delta (R,t=0)=\partial_t\delta
(R,t)|_{t=0}$.
 Thence the
 integration of (\ref{15}) along $\tilde \Gamma_{a, (R,t)}$ yields
\begin{equation}
 H_M\le \int_a^{\infty }{dr \over\eta_r}|RHS(r)|,
\end{equation}
where $RHS(r)$ stands for the right hand side of Eq. (\ref{15}).
Our task consists in estimating the line integral of $|RHS(r)|$.

In order to do this one uses the  estimates of  (\ref{7}). The
 calculation is quite long and
we will describe only the main points. In the first step one uses the
Schwarz
inequality in the right hand side of (\ref{15}), in order to obtain
an expression of the type
\begin{eqnarray}
4m\Biggl( \int_R^{\infty }dr {(\partial_t\delta )^2\over \eta_r }
\Biggr)^{1/2}
\times  \Biggl( \int_R^{\infty }dr \eta_r{f^2(r)\over r^8}\Biggr)^{1/2},
\label{15a}
\end{eqnarray}
where $f(r)$ denotes  $ (-3 \Psi_1)^2$, $(2\Psi_2 / r)^2$, or the squares
of the terms that are proportional to $63m$. The   first integral
can be bounded by $\sqrt{H(R)}$, therefore  (\ref{15}) and
(\ref{15a})  yield
\begin{eqnarray}
&&
(\partial_t+\partial_{r^*})\sqrt{H(R,t)}
 \le
2m\Biggl( \int_R^{\infty }dr\eta_r {9\Psi_1^2\over r^8}\Biggr)^{1/2}+
2m\Biggl( \int_R^{\infty }dr{4\Psi_2^2\over r^{10}} \Biggr)^{1/2}
+\nonumber\\
&&2m\Biggl( \int_R^{\infty }dr\Bigl({ 63m (1+{m\over r})
 \over 4 (1+{3m\over 2r})^2}\Bigr)^2
  {\Psi_0^2\over r^8}\Biggr)^{1/2}+
2m\Biggl( \int_R^{\infty }dr\Bigl({ 63m (1+{m\over r})
 \over 4 (1+{3m\over 2r})^2}\Bigr)^2
  {\Psi_1^2\over r^{10}}\Biggr)^{1/2}+
\nonumber\\
&&2m\Biggl( \int_R^{\infty }dr\Bigl({ 63m (1+{m\over r})
 \over 4 (1+{3m\over 2r})^2}\Bigr)^2
  {\Psi_2^2\over r^{12}}\Biggr)^{1/2} .
\label{15b}
\end{eqnarray}
The integrands of (\ref{15b}) are taken at a time $t$ and
$(R,t)\in \tilde \Gamma_{a}$ ; the integration
extends over the part $r\ge R$ of the Cauchy hypersurface $\Sigma_t $.
At this place one inserts the bounds on $\Psi_0$, $\Psi_1$ and $\Psi_2$.
That requires  some care; the estimates hold true on the initial
hypersurface $\Sigma_0 $, while here one needs estimates on $\Sigma_t $.
This point is clarified later.
It is useful to introduce dimensionless variables $x=R/a$ and $\tilde m=
m/a$.

i) First we shall consider the contribution that is due to $3\Psi_1$.
Let $r_0$ be defined by $(r,t)\in   \tilde \Gamma_{r_0}$.
 The insertion of the bound given in (\ref{7})  bounds
 $2m\Biggl( \int_R^{\infty }dr\eta_r {9\Psi_1^2\over r^8}\Biggr)^{1/2}$ by
$ 6mC_1\Biggl( \int_R^{\infty }dr\eta_r^4
 {g_1(r_0)\over r^5}\Biggr)^{1/2}$. Notice that
  $g_1(r)$ is an increasing function,
  therefore  if one replaces $g_1(r_0) $ by $g_1(r)$, then the integral
that appears here cannot be smaller.
 In this way one  utilizes the initial information
(the energy inequality (\ref{7})) in order to control the evolution.
 The integral in question  can be
performed explicitly, with the result
\begin{eqnarray}
 6mC_1\Biggl( \int_R^{\infty }dr\eta_r^4
 {g_1(r)\over r^5}\Biggr)^{1/2}
 ={6mC_1\over a^2}\sqrt{G_1(x)}.
\label{15c}
\end{eqnarray}
Here,  it holds
\begin{eqnarray}
&&- G_1(x) =
{\tilde m^4 (137 - 770 \tilde m + 1880 \tilde m^2 - 2160 \tilde m^3 +
960\tilde m^4  )
  \over 30 (-1 + \tilde 2m)^5 x^8} -\nonumber\\
&&  {\tilde m^3 (991 - 5110 m + 10840 \tilde m^2 - 8880 \tilde m^3 -
720\tilde m^4 +
 3360\tilde m^5 )
   \over 105 (-1 + 2 m)^5 x^7} +\nonumber\\
&& {\tilde m^2    (2981 - 13010\tilde m + 18440\tilde m^2 + 7920\tilde m^3
- 41520\tilde m^4
+ 27360\tilde m^5)
  \over  420(-1 + 2\tilde m)^5 x^6} -\nonumber\\
 &&  {\tilde m  (4497 - 11370\tilde m - 21720\tilde m^2 + 133040\tilde m^3
- 200240\tilde m^4 +
101600\tilde m^5)
 \over 2100 (-1 + 2\tilde m)^5 x^5}+\nonumber\\
&&  {375 + 4650\tilde m - 35400\tilde m^2 + 93200\tilde m^3 -
   110000\tilde m^4 + 49376\tilde m^5  \over 3360 (-1 + 2\tilde m)^5
x^4}-\nonumber\\
&&  {11\over 336\tilde m x^3} - {11\over 448\tilde m^2 x^2}
 - {11\over 448\tilde m^3 x} +   {11 \ln({x\over -2\tilde m + x}) \over
896\tilde m^4}
-\nonumber\\
 &&  ( 280\tilde m^4 - 640\tilde m^3x + 560\tilde m^2 x^2 - 224\tilde m
x^3 + 35 x^4 )
    {\ln ({-2\tilde m + x\over 1 - 2\tilde m}\over 140 x^8}).
\label{15d}
\end{eqnarray}
This rather long expression is quite well approximated by  $ G_1=
 (1+ 4\ln x  )/(16x^4)$
if $m/a<<1$.  The integration of  (\ref{15c}) along a null cone $C_{a}$
is  done  as follows.
The integral $\int_1^{\infty }\sqrt{G_1(x)} $ is bounded from above,
\begin{eqnarray}
\int_1^{\infty }dx\eta_x^{-1}\sqrt{G_1(x)}\le
\Biggl( \int_1^{\infty }dxx^2\ G_1(x) \Biggr)^{1/2}
\Biggl( \int_1^{\infty }dxx^{-2}\eta_x^{-2}  \Biggr)^{1/2}.
\end{eqnarray}
Numerical integration yields
\begin{eqnarray}
&&6C_1{m\over a}\sqrt{\int_1^{\infty }{(1+ 4\ln x )\over 16x^2 }}
+O((m/a)^2)
\approx  \nonumber\\
&&  8.24   {m\sqrt{E(a,0) }\over a}  +O((m/a)^2).
\label{15d1}
\end{eqnarray}
One can check that the neglected terms can give a   contribution
comparable
to the leading term only at distances smaller than $6.6m$.

ii) The calculation concerning the contribution of the $\Psi_2$ function
is similar.  The leading (proportional to $m^0$) term is $\int_R^{\infty }
 dr{\Psi_2^2\over r^{10}}$. $|\Psi_2|$ is bounded in terms of  $g_2(x)$.
$g_2(x)$ is an increasing function,  and a reasoning  similar
to what was made when discussing $g_1(x)$, leads to the conclusion that
one can again use the initial energy inequality given by   (\ref{7}).
  One finds that   $\int_R^{\infty } dr{\Psi_2^2\over r^{10}}$
is bounded from above by
\begin{eqnarray}
&&-{1\over a^4} G_2(x):=-{1\over a^4}\int_x^{\infty }dy{g_2(y)\over y^6}
(1-2\tilde m/y)^2=
\nonumber \\
&&{4\tilde m^2(-3 + 44\tilde m - 120\tilde m^2 + 96\tilde m^3)\over
21(-1+ 2\tilde m)^3x^7} - \nonumber\\
&& {2\tilde m(-21 + 236\tilde m - 408\tilde m^2 - 192 m^3 + 576\tilde m^4)
\over 63(-1 + 2\tilde m)^3x^6} + \nonumber \\
&& {-21 + 88\tilde m + 480\tilde m^2 - 1968\tilde m^3 + 1760\tilde m^4
\over 105 (-1 + 2\tilde m)^3\tilde x^5} - \nonumber\\
&& {34\over 105 x^4} - {31\over 630\tilde mx^3} -
{31\over 840\tilde m^2x^2}  - \nonumber \\
&&  {31\over 840\tilde m^3x} +
 {31\ln ({x\over x-2\tilde m})\over 1680\tilde m^4}-\nonumber\\
&& {8\tilde m(60\tilde m^2 - 70\tilde mx +
21x^2)\ln ({-2\tilde m + x\over 1 - 2m})\over 105x^7}.
\label{15e}
\end{eqnarray}
In the limit of $m\rightarrow 0$ the function $G_2(x)$ coincides with
$(-4+5x)/(20x^5)$. Similarly as before, in order to get a term
bounding $\sqrt{H_M}$, one should integrate
$\sqrt{G_2(x)}/(1-2\tilde m/x)$ along a  null cone $C_{a}$. That gives
0.15, up to terms $O(m)$, after
manipulations similar to those done earlier. The $O(m)$
correction becomes dominant when  $2m/a> 0.3$.
After a reasoning similar to that applied   above in the case of $\Psi_1$
one finds that  the total contribution due to the bound on the
$\Psi_2 $ function is equal to
\begin{equation}
4C_2\sqrt{\int_1^{\infty }dx {(-4+5x)\over 20x^3 }} {m\sqrt{E(a,0) }\over
a}
=2.19{m\sqrt{E(a,0) }\over a}.
\end{equation}
In summary, one obtains
\begin{equation}
\sqrt{H_M} \le 10.43{m\sqrt{E(a,0) }\over a}+O(m^2).
\label{16}
\end{equation}
In the above analysis, in (\ref{15b}), we neglected the terms
proportional to $63m$. They give corrections of the order $O(m^2)$
to the right hand side of (\ref{16}).
 We checked that their contribution is  small in the region $a>6.6m$.
Our final result (\ref{27}) tells us that $a>\sqrt{218}m \approx 15m$
is valid for a nontrivial estimate. Therefore, all higher order
terms in (\ref{16}) can be safely neglected.

\end{document}